\documentclass[9pt,twocolumn]{IEEEtran}
\usepackage[ruled, linesnumbered]{algorithm2e} 
\usepackage[utf8]{inputenc} 
\usepackage[T1]{fontenc} 
\usepackage{amsmath, amssymb, amsfonts} 
\usepackage{graphicx} 
\usepackage{subcaption} 
\usepackage[font=footnotesize]{caption} 
\usepackage{booktabs, array, tabularx, multirow} 
\usepackage{float} 
\usepackage{enumitem} 
\usepackage[colorlinks=true, allcolors=blue]{hyperref} 
\usepackage[margin=0.75in]{geometry} 
\usepackage[ruled, linesnumbered]{algorithm2e} 
\usepackage{amsmath, amssymb, amsfonts} 
\usepackage{cite} 
\usepackage{amsmath, amssymb, amsfonts}
\begin{document}

\title{Optimizing CDN Architectures: Multi-Metric Algorithmic Breakthroughs for Edge and Distributed Performance}

\author{%
\IEEEauthorblockN{Md Nurul Absur\textsuperscript{*}, Sourya Saha\textsuperscript{*}, Sifat Nawrin Nova\textsuperscript{\dag}, Kazi Fahim Ahmad Nasif\textsuperscript{\ddag}, Md Rahat Ul Nasib\textsuperscript{\S}}\\
\IEEEauthorblockA{\textsuperscript{*}Department of Computer Science, City University of New York, New York, USA\\
\textsuperscript{\dag}Department of Computer Science, Chalmers University of Technology, Gothenburg, Sweden\\
\textsuperscript{\ddag}College of Computing and Software Engineering, Kennesaw State University, Georgia, USA\\
\textsuperscript{\S}Samsung Austin Research Center, Austin, Texas, USA}\\
\IEEEauthorblockA{Emails: mabsur@gradcenter.cuny.edu, ssah42@gradcenter.cuny.edu, esifatn@chalmers.se, knasif@students.kennesaw.edu, nasib131@gmail.com}
}

\maketitle

\thispagestyle{empty} 
\vspace{-10pt}
\begin{abstract}

A Content Delivery Network (CDN) is a powerful system of distributed caching servers that aims to accelerate content delivery, like high-definition video, IoT applications, and ultra-low-latency services, efficiently and with fast velocity. This has become of paramount importance in the post-pandemic era. Challenges arise when exponential content volume growth and scalability across different geographic locations are required. This paper investigates data-driven evaluations of CDN algorithms in dynamic server selection for latency reduction, bandwidth throttling for efficient resource management, real-time Round Trip Time analysis for adaptive routing, and programmatic network delay simulation to emulate various conditions. Key performance metrics, such as round-trip time (RTT) and CPU usage, are carefully analyzed to evaluate scalability and algorithmic efficiency through two experimental setups: a constrained edge-like local system and a scalable FABRIC testbed. The statistical validation of RTT trends, alongside CPU utilization, is presented in the results. The optimization process reveals significant trade-offs between scalability and resource consumption, providing actionable insights for effectively deploying and enhancing CDN algorithms in edge and distributed computing environments.

\end{abstract}


\section{Introduction}

The importance of video streaming has grown significantly in the digital age, becoming a dominant force in internet traffic and user engagement. Recent statistics indicate that video accounts for over 82\% of the traffic on the internet \cite{10.1016/j.jnca.2024.103889}. The shift to remote work, virtual learning, and digital entertainment has not only increased this demand. Still, it has also created an urgent necessity to address the challenges of delivering high-quality video content reliably and at scale. This necessitates innovative network management solutions to accommodate the growing and diverse user demands.

The unprecedented data volume and complexity growth presents significant challenges for existing Content Delivery Networks (CDNs). Traditional CDN architectures, optimized for static and moderately dynamic content, are increasingly strained by the demands of real-time, ultra-high-definition video and other bandwidth-intensive applications. Achieving scalability across geographically dispersed servers and efficiently handling dynamic content remain critical bottlenecks \cite{PLAGEMANN2006551}. To ensure an optimal Quality of Experience (QoE), modern CDNs must address several urgent issues: managing the surging content volume, ensuring seamless scalability across diverse locations, and delivering ultra-low latency under varying network conditions \cite{10617319}. Additionally, the shift toward localized and personalized content and the growing prevalence of interactive and immersive experiences necessitates innovative CDN design and management approaches \cite{MA2020511}. {\em Thus, a scalable and reliable CDN configuration scheme is needed to consider advanced dynamic server selection, bandwidth throttling, delay modifications in the presence of essential performance metrics, and advanced statistical multi-metric decision outcomes. }

Our contribution addresses the current limitations of Content Delivery Networks (CDNs) by introducing dynamic algorithms and a comprehensive multi-metric analysis framework. These advancements aim to improve video streaming performance by reducing latency, enhancing adaptive streaming capabilities, and optimizing resource utilization across distributed systems. Our work focuses on key challenges in CDN performance management, ensuring better scalability, load balancing, and service quality. The key innovations of our approach include:

\begin{itemize}
    \item The development of dynamic multi-algorithm frameworks aims to optimize the performance of Content Delivery Networks (CDNs).
    \item Implement programmatic modifications to network delays in order to simulate edge and distributed environments for scalability testing.
    \item A multi-metric statistical analysis aims to overcome the limitations of performance management that rely on a single metric.
    \item The dataset, acquired from real-life simulations, is shared for open use to foster research and development.
\end{itemize}

The rest of the paper is organized as follows. Section~\ref{sec: related work} discusses the related work in this area. Section~\ref{sec:systemmodel} introduces our solution approach with different configurations, and Section~\ref{sec:experimentresults} discusses performance evaluation. Finally, Section~\ref{sec:conclusions} concludes the paper addressing future works.
\color{black}
\section{Related Works}
\label{sec: related work}

Content Delivery Network (CDN) optimization is an active area of research that has garnered significant attention in recent years. The applications of CDN are pretty diverse, and the variability of CDN management has made this use case ubiquitous in terms of system optimization, network control, resource efficiency, and so on. Software-Defined Networking (SDN) is gaining substantial attention to address scalability and cost challenges in Content Delivery Networks (CDNs). Yang et al. \cite{10103706} have proposed a Software-Defined Content Distributed Network (SDCDN) solution. Another alternative model is implemented by combining Peer-to-Peer (P2P) and CDN together to achieve optimal content experience \cite{5764567}. Despite this, the flexibility and global reach offered by CDNs continue to make them the preferred choice for video streaming and
real-time multimedia services \cite{10.1145/3625468.3652196}.

To address the shortcomings of CDNs, Zhou et al. use a load balancing algorithm based on playback volume and the remaining resources of the server to reduce latency and enhance Quality of Experience (QoE) \cite{10507423}. Vaton et al. measure web browsing performance using different CDN to decrease loading times by up to 400 percent \cite{9040260}. In \cite{10622908}, a new anycast distribution method is proposed that incorporates spatial locality of the popularity of the content and addresses large-scale scalability. Recent progress has also centered on refining content delivery networks (CDNs) by implementing P2P-CDN architecture \cite{10330627} and integrating multi-source streaming to provide an exceptional viewing experience \cite{9842378}.

Working on a CDN in a testbed environment aims to replicate a scalable, twin-like large network system and measure network variability. This approach is becoming popular in the video content and streaming research community. In \cite{9615676}, the authors introduced a CDN emulator to carry out discrete event
simulation and considered resource consumption a performance metric. The FABRIC testbed is an innovative
and rapidly emerging platform that comprises a wide
range of use cases and significant computational capabilities \cite{8972790}. This infrastructure offers diverse computational
resources, and researchers are leveraging its power to
address many complex network management challenges
through various applications \cite{9829810}\cite{10620897}. Some works use machine learning and statistical methods to make CDNs more ubiquitous, better performing, scalable, and applicable to the industry \cite{9155373}, \cite{US11659015B2}.

\section{System Model \& Problem Formulation}
\label{sec:systemmodel}

To tackle CDN architecture, we formulated problem statements and sought optimal solutions by implementing two types of configurations. The first configuration consisted of a large-scale distributed CDN server setup, which emulated high-end computation and scalability within the FABRIC testbed. The second configuration featured an edge device with minimal system requirements, representing a small CDN setup that can still be utilized in various applications to address cost constraints.

\subsection{Testbed Setup For Scalable Implementation}

\begin{figure*}[htbp]
    \centering

    \begin{subfigure}{0.37\textwidth} 
        \centering
        \includegraphics[width=\linewidth, height=0.7\linewidth]{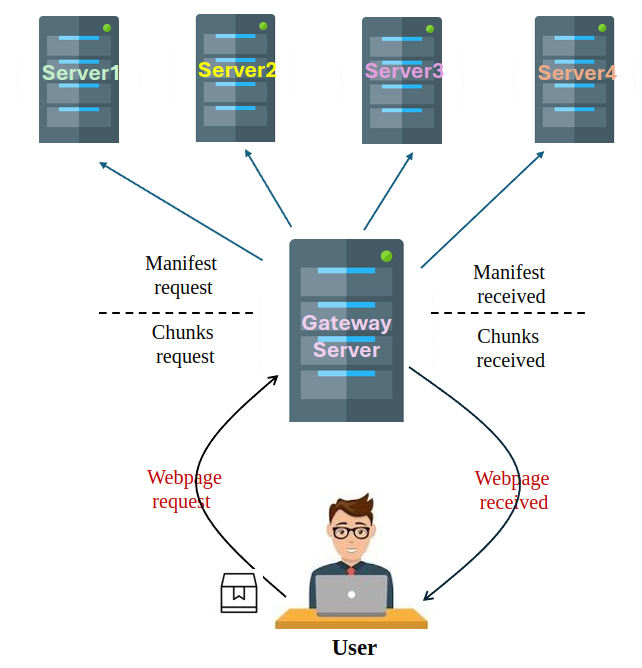}  
        \caption{Topology Overview of Testbed CDN Architecture} 
        \label{fig:cdntestbed} 
    \end{subfigure}
    \hfill
    \begin{subfigure}{0.35\textwidth} 
        \centering
        \includegraphics[width=\linewidth, height=0.7\linewidth]{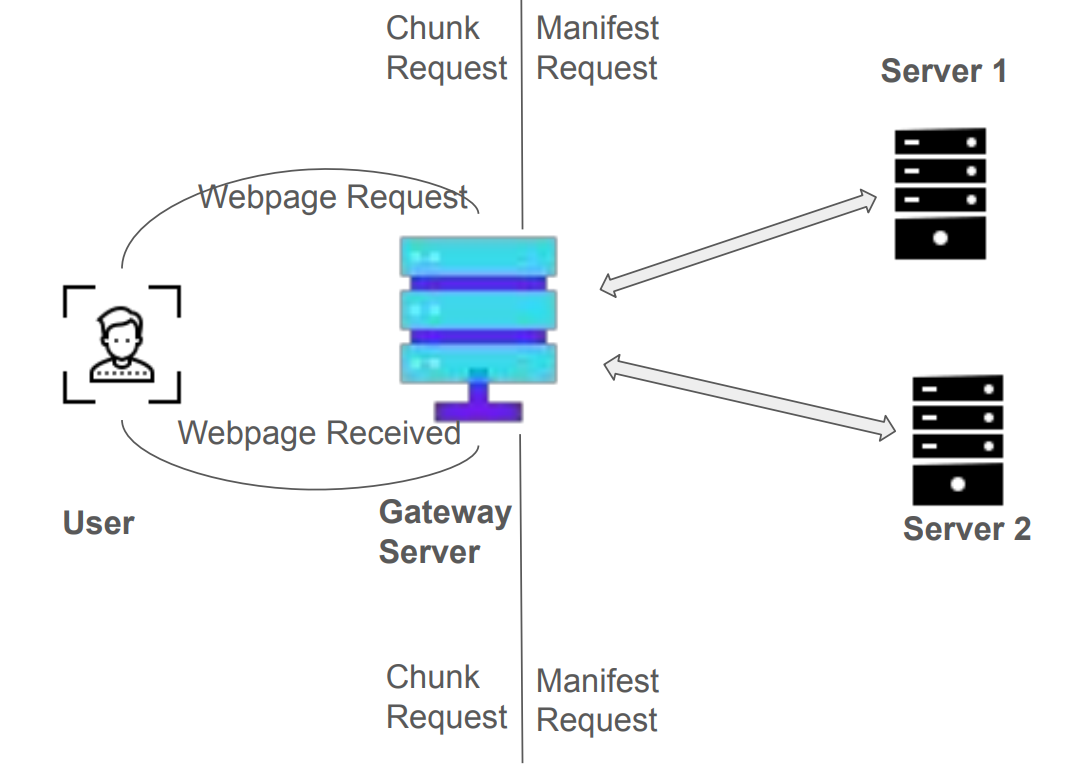}  
        \caption{Local Edge CDN Architecture} 
        \label{fig:edgeCDN} 
    \end{subfigure}

    \caption{CDN Setup Across Different Configurations}
    \label{fig:MeanSDReliabilityForTwoDatasets}
\end{figure*}

The architecture of our testbed Content Delivery Network (CDN) implementation, as illustrated in Figure of \ref{fig:cdntestbed}, underscores the pivotal role of distributed servers—Server1, Server2, Server3, and Server4—in facilitating the communication process. Central to this architecture is the Gateway Server, which orchestrates interactions between the user and the distributed servers. When a user requests a webpage, the initial query is directed to the Gateway Server, which issues manifest requests to the various distributed servers. The manifest, containing essential metadata regarding the content chunks stored across these servers, enables the Gateway Server to determine the appropriate servers for subsequent chunk requests efficiently. These content chunks, representing optimized segments of the webpage, are then retrieved and delivered to the user by the designated servers. This streamlined process maximizes throughput and minimizes latency, exemplifying the scalability of our CDN model, which is designed to emulate real-world, large-scale server distribution across diverse geographical locations.

\subsection{Local Edge-like System Setup}

In Figure \ref{fig:edgeCDN}, a CDN architecture is deployed on local devices to emulate edge systems, enabling performance evaluation under dynamic, real-time scenarios. Upon a user’s webpage request, the Gateway Server initiates manifest requests to multiple servers (e.g., Server 1 and Server 2), which respond with the necessary metadata. The Gateway Server then fetches specific content chunks, ensuring efficient delivery while minimizing latency and enhancing user experience. To optimize performance further, the system diligently monitors round-trip time (RTT) values, dynamically selecting servers with lower RTTs to maintain high reliability and throughput, and effectively simulates real-world CDN operations.

\subsection{Advance CDN Setup Algortihms}

Four advanced algorithms are used to make the overall CDN setup more dynamic, adaptable to throttled segmentation, and to address network delay issues. These algorithms measure CDN performance using various metrics and establish which setup is better suited for scalability.

Algorithm \ref{alg:DASH_HTTP_Redirects} demonstrates an implementation for streaming video using the DASH protocol while handling HTTP redirects. It initializes a DASH player, intercepts HTTP requests to process media segments dynamically, constructs redirect URLs, and updates the requests accordingly, all while logging essential details.

\begin{algorithm}
\DontPrintSemicolon
\caption{DASH Video Streaming with HTTP Redirects}
\label{alg:DASH_HTTP_Redirects}
\KwData{\textit{videoElement}, \textit{initialMpdUrl}}
\KwResult{Optimized media segment handling}

\SetKwProg{Fn}{Function}{:}{}
\Fn{DASHStreaming(videoElement, initialMpdUrl)}{
    $player \gets$ InitializePlayer(videoElement, initialMpdUrl)\;
    OverrideXMLHttpRequest()\;
    $player$.\text{on}(``\texttt{FRAGMENT\_LOADING}'', ProcessSegment)\;
}

\Fn{ProcessSegment(event)}{
    \If{IsMediaSegment(event.request)}{
        $fileName \gets$ ExtractFileName(event.request.url)\;
        $redirectUrl \gets$ ConstructRedirectUrl(fileName)\;
        UpdateRequest(event.request, redirectUrl)\;
        LogRequest(fileName)\;
    }
}

\Fn{InitializePlayer(video, url)}{\KwRet new Player(video, url)\;}
\Fn{OverrideXMLHttpRequest()}{
    Intercept HTTP requests to log and modify as needed\;}
\Fn{IsMediaSegment(request)}{\KwRet request.url.endswith(".mp4")\;}
\Fn{ExtractFileName(url)}{\KwRet url.split("/")[-1]\;}
\Fn{ConstructRedirectUrl(fileName)}{\KwRet \texttt{http://cdn.example.com/} + fileName\;}
\Fn{UpdateRequest(request, url)}{request.url $\gets$ url\;}
\Fn{LogRequest(fileName)}{Output log information about the request\;}

\end{algorithm}

Algorithm \ref{algo:MPD} processes an MPD file from a specified URL, retrieves metadata such as the video ID and segment count, and downloads the media segments while implementing throttling to mimic real-world network conditions. The utility function handles file writing correctly while adhering to throttling restrictions.

\begin{algorithm}
\DontPrintSemicolon
\caption{Download and Parse MPD with Throttled Segment Download}
\label{algo:MPD}
\KwData{\texttt{php\_url} (PHP script URL)}
\KwResult{Total media segments downloaded}

Parse MPD from \texttt{php\_url}; extract video ID, segment count\;
Download initial segment\;
\For{each segment in MPD}{Construct URL, download with throttle\;}
\Return{Number of segments}\;

\SetKwProg{Fn}{Function}{:}{end}
\Fn{\FuncSty{save\_and\_throttle\_download}{response, filename}}{
    Write response to file with throttle\;
}
\end{algorithm}

Algorithm \ref{algo:ping RTT} pings a list of servers, measures the round-trip time (RTT) for each server, and selects the one with the lowest RTT for directing clients. If no valid RTT is found, a random server will be chosen as a fallback option.

\begin{algorithm}
\DontPrintSemicolon
\caption{Select Server Based on Ping RTT}
\label{algo:ping RTT}
\KwResult{Redirect to server with the lowest RTT}

Initialize min RTT to $\infty$, selected server to null\;
\SetKwProg{Fn}{Function}{:}{end}
\Fn{\FuncSty{getPingRTT}(serverIP)}{
    Ping server, extract RTT\; 
    \KwRet RTT or high value on failure\;
}

\ForEach{server in list}{
    $RTT \gets$ \FuncSty{getPingRTT}(Extract IP(server URL))\;
    \If{$RTT <$ min RTT}{min RTT $\gets$ RTT; update selected server\;}
}

\If{no valid RTT}{Select random server\;}

Redirect client to selected server with video filename\;
\end{algorithm}

Algorithm \ref{algo:linux} adjusts network delay using Linux Traffic Control by introducing a random delay between 200 and 800 milliseconds in a continuous loop. It verifies these changes and pauses for a duration determined by a Poisson distribution.

\begin{algorithm}
\DontPrintSemicolon
\caption{Dynamic Network Delay Modification with Linux Traffic Control}
\label{algo:linux}
\KwResult{Continuously updated network delay}

\While{True}{
    Generate random delay (200-800ms)\;
    Remove existing delay; apply new delay\;
    Verify and print success of removal/addition\;
    Set and print sleep time (Poisson distribution)\;
    Sleep for the calculated duration\;
}
\end{algorithm}

\subsection{Dataset Collection For Multimetric Correlation Analysis}

Relying on a single metric is insufficient for making scalability decisions in \emph{CDN}. To support this, we gathered extensive data using various real-life mimicking configurations: \emph{four servers, eight servers, and twelve servers}. We collect a vast number of \emph{RTT} \& \emph{CPU consumption} data to find a pattern and make trade-offs between cost management and scalability decisions. This dataset can be found at \texttt{https://github.com/Tomxx7/FABRIC\_RTT\_CPU.git}

Algorithm \ref{algo:ping_rtt_recording} pings a list of servers multiple times (\texttt{num\_pings}) to measure \textbf{Round Trip Time (RTT)}, modifies the RTT with a Poisson-distributed random value, and logs the results (including a timestamp) to a CSV file (\texttt{ping\_results1000.csv}). It captures RTT for each server in real-time, handles ping failures gracefully, and delays each round by one second.

\begin{algorithm}
\caption{Ping Servers and Record RTT with Poisson Modification}
\label{algo:ping_rtt_recording}

\KwData{\texttt{servers}, \texttt{num\_pings}, \texttt{output\_file}}
\KwResult{CSV file with timestamps and RTT for each server}

\SetKwProg{Fn}{Function}{:}{}
\SetKwFunction{Ping}{ping\_server}
\SetKwFunction{Record}{record\_results}

\Fn{\Ping{\texttt{ip\_address}}}{
    Execute \texttt{ping} command and extract RTT\;
    Add Poisson random value (\texttt{mean}, $\lambda = 200$ms)\;
    \Return Modified RTT or \texttt{None}\;
}

\Fn{\Record{\texttt{servers}, \texttt{output\_file}}}{
    Write headers (\texttt{servers}) to \texttt{output\_file}\;
    \For{$i \gets 1$ to \texttt{num\_pings}}{
        Initialize \texttt{row} with timestamp\;
        \ForEach{\texttt{server} in \texttt{servers}}{
            Append RTT or \texttt{N/A} to \texttt{row}\;
        }
        Write \texttt{row} to \texttt{output\_file}; Wait 1 second\;
    }
}

\Record{\texttt{servers}, \texttt{output\_file}}
\end{algorithm}

Algorithm \ref{algo:cpu_utilization}  fetches \textbf{CPU utilization} data from Prometheus by querying total and active CPU times, calculates the utilization percentage, and adds a Poisson-distributed random value to simulate variability. The data and timestamp are logged into a CSV file every 2 seconds for 1500 iterations. It handles errors gracefully and ensures missing values are marked as "N/A" in the output.  Prometheus is an open-source, research-driven monitoring system optimized for time-series data collection and analysis in distributed and cloud-native environments \cite{prometheus2024}. It provides a robust pull-based metric scraping mechanism. It leverages PromQL, a powerful query language, to enable advanced metric correlations and real-time insights, supporting scalability and precision in dynamic systems.

\begin{algorithm}
\DontPrintSemicolon
\caption{CPU Utilization Logging with Prometheus}
\label{algo:cpu_utilization}

\KwData{\texttt{PROMETHEUS\_URL}, \texttt{TOTAL\_CPU\_QUERY}, \texttt{ACTIVE\_CPU\_QUERY}}
\KwResult{CSV file of CPU utilization percentages}

\SetKwProg{Fn}{Function}{:}{}
\Fn{fetch\_cpu\_utilization()}{
    Query Prometheus for total and active CPU times\;
    Compute utilization: $\frac{\mathrm{Active\_CPU}}{\mathrm{Total\_CPU}} \times 100$\;
    Add Poisson noise ($\lambda = 30$)\;
    \Return utilization per instance\;
}

\Fn{write\_to\_csv(writer, data, timestamp)}{
    Initialize header if first write\;
    Write \texttt{timestamp} and utilization (or \texttt{N/A} for missing values)\;
}

\textbf{Main Procedure:}\;
Open CSV file\;
\For{$i \gets 1$ to 1500}{
    Record \texttt{current\_time} and fetch utilizations\;
    Write data to CSV\;
    Sleep for 2 seconds\;
}
\end{algorithm}

\section{Experiment Results}
\label{sec:experimentresults}

\begin{figure*}[htbp]
    \centering

    \begin{subfigure}{0.22\textwidth}
        \centering
        \includegraphics[width=\linewidth, height=1.1\linewidth]{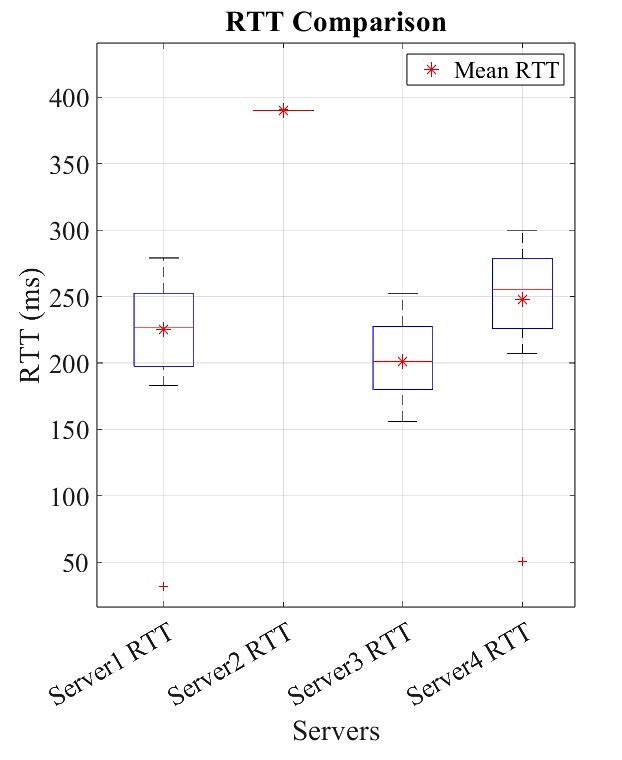}  
        \caption{RTT Distribution on FABRIC}
        \label{fig:RTTdisFABRIC}
    \end{subfigure}
    \hfill
    \begin{subfigure}{0.22\textwidth}
        \centering
        \includegraphics[width=\linewidth, height=1.1\linewidth]{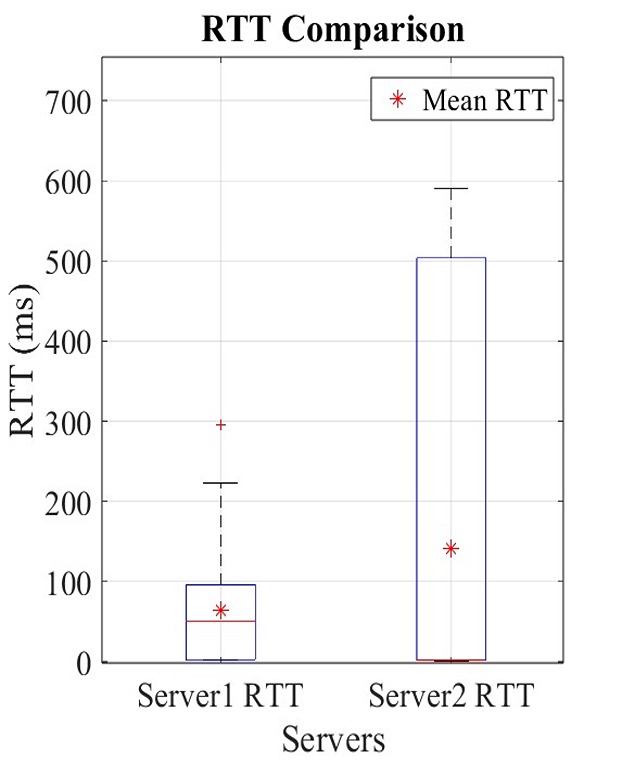}  
        \caption{RTT Distribution on Edge}
        \label{fig:RTTdisedge}
    \end{subfigure}
    \hfill
    \begin{subfigure}{0.22\textwidth}
        \centering
        \includegraphics[width=\linewidth, height=1\linewidth]{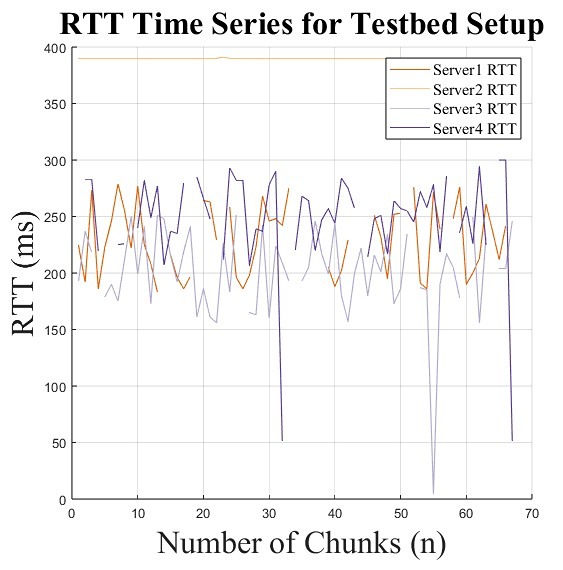}  
        \caption{Time Series Distribution on FABRIC}
        \label{fig:timeseriesRTTFabric}
    \end{subfigure}
    \hfill
    \begin{subfigure}{0.22\textwidth}
        \centering
        \includegraphics[width=\linewidth, height=1.1\linewidth]{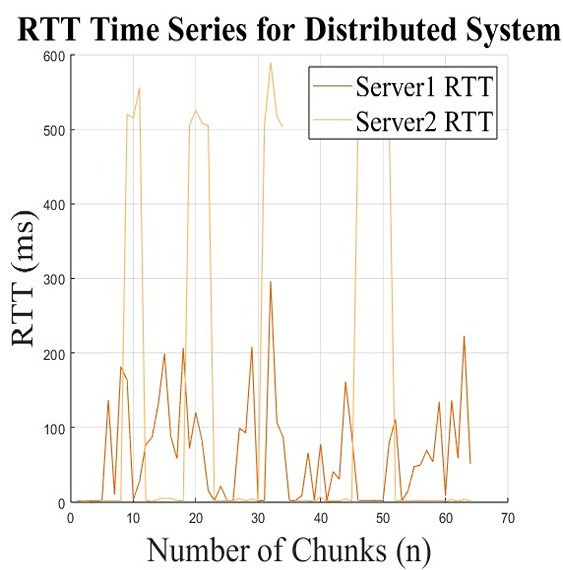}  
        \caption{Time Series Distribution on Edge}
        \label{fig:timeseriesRTTedge}
    \end{subfigure}

    \caption{Performance Analysis Based On RTT Across Setups}
    \label{fig:PerformanceRTTAnalysis}
\end{figure*}

\subsection{Scalability}

To explore more effective scalability setups for a CDN system, we examine four distinct servers in the FABRIC testbed that simulate large-scale CDN servers, along with two servers intended for a smaller edge-like CDN configuration.

In Figure \ref{fig:PerformanceRTTAnalysis}, we evaluate the Round-Trip Time (RTT) performance across two distinct setups. This analysis will feature two visualizations: a boxplot that illustrates the distribution and variability of RTT values and a time series distribution that highlights factors such as network congestion, hardware limitations, and geographical distances between nodes. These visualizations will provide valuable insights into the latency performance of each server, aiding in the optimization of server selection for improved network efficiency and reduced latency.

Figure \ref{fig:RTTdisedge} illustrates the inherent challenges associated with edge-based or small-scale CDN setups, where Server 2 consistently demonstrates high variability and instability in round-trip time (RTT), significantly affecting performance and user experience. Even within the more controlled FABRIC testbed shown in Figure \ref{fig:RTTdisFABRIC}, which includes four servers, Server 2 fails, resulting in an infinite RTT that indicates its inability to respond. This observation reveals a critical limitation: even a single point of failure can jeopardize overall service reliability despite the presence of multiple servers. While edge setups may present cost advantages, their lack of robustness and greater susceptibility to failures render them less suitable for real-world applications that require uninterrupted and reliable content delivery. These findings emphasize prioritizing resilience and stability in CDN architectures to ensure seamless user experiences.

Figures \ref{fig:timeseriesRTTFabric} and \ref{fig:timeseriesRTTedge} illustrate a comparative Round Trip Time (RTT) analysis between Testbed and Distributed System configurations, highlighting their respective advantages and limitations. The Testbed setup demonstrates stable RTT across four servers, with Server 2's absence resulting in infinite RTT, highlighting vulnerabilities even in controlled environments. This setup is ideal for high-reliability applications like finance and healthcare. In contrast, the edge setup shows significant RTT variability, particularly with Server 2, which is affected by geographic and traffic fluctuations. While this variability challenges latency-sensitive applications, it offers valuable insights for stress-testing CDNs, helping identify failure points and optimization strategies. The Testbed ensures consistency, while the Distributed System provides crucial data for improving CDN scalability and resilience.

\subsection{Multi Metric Trade-off Analysis}

Relying on a single metric, such as round-trip time (RTT) or CPU utilization, can lead to misleading conclusions about system performance and scalability. For instance, while CPU utilization may indicate a stable average of around 30\%, it can obscure increasing RTT, which points to rising latency as server count increases. Conversely, RTT highlights scalability issues due to coordination delays but does not specify whether computational bottlenecks or network inefficiencies cause these. Analyzing both metrics reveals that although CPU load may stabilize with more servers, communication overhead significantly hampers network performance. This comprehensive approach is vital for identifying the actual limitations of scalability and opportunities for optimization.

\begin{figure*}[htbp]
    \centering

    \begin{subfigure}{0.22\textwidth}
        \centering
        \includegraphics[width=\linewidth, height=1.1\linewidth]{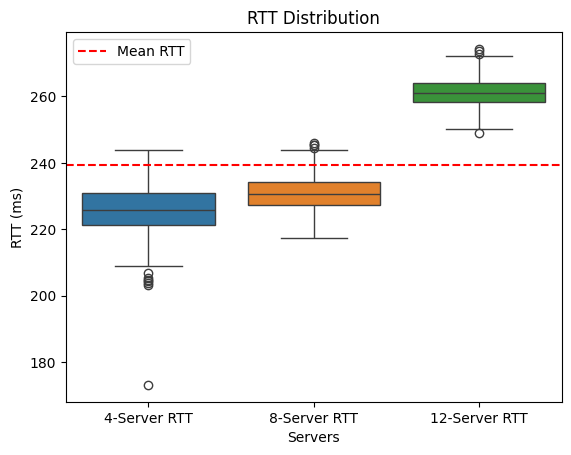}  
        \caption{RTT Distribution Across Configurations}
        \label{fig:multimeanrtt}
    \end{subfigure}
    \hfill
    \begin{subfigure}{0.22\textwidth}
        \centering
        \includegraphics[width=\linewidth, height=1.1\linewidth]{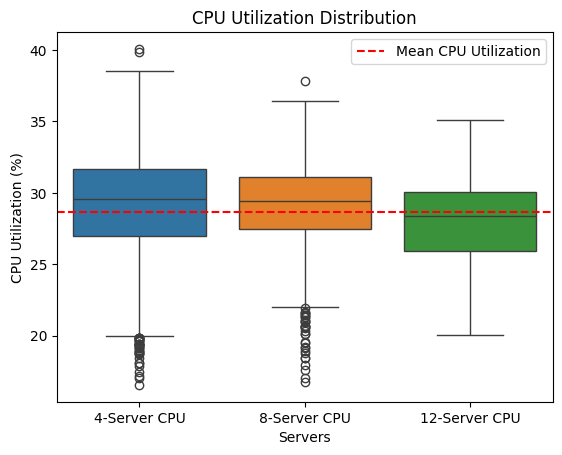}  
        \caption{CPU Utilization Distribution Across Configurations}
        \label{fig:multimeancpu}
    \end{subfigure}
    \hfill
    \begin{subfigure}{0.22\textwidth}
        \centering
        \includegraphics[width=\linewidth, height=1.1\linewidth]{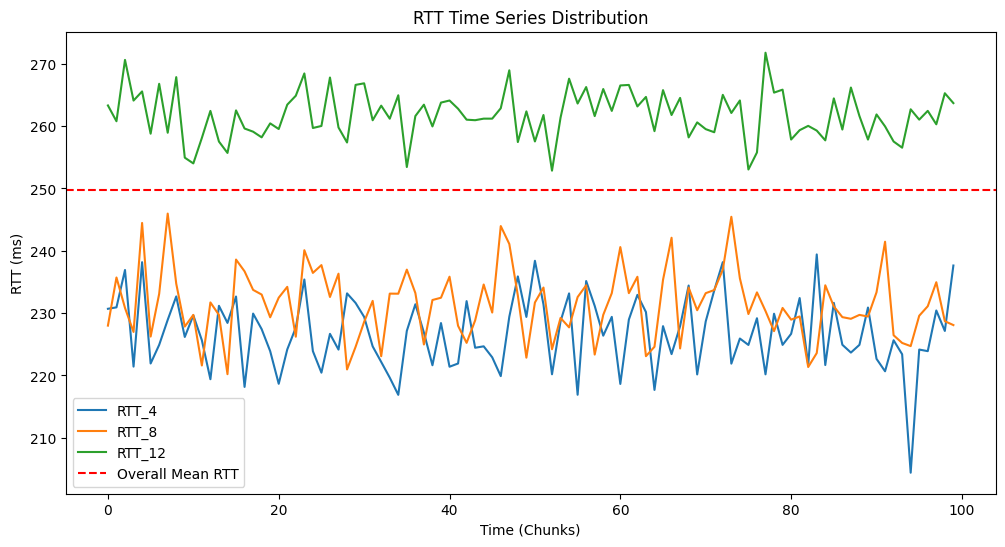}  
        \caption{RTT Time Series Across Configurations}
        \label{fig:multitimertt}
    \end{subfigure}
    \hfill
    \begin{subfigure}{0.22\textwidth}
        \centering
        \includegraphics[width=\linewidth, height=1.1\linewidth]{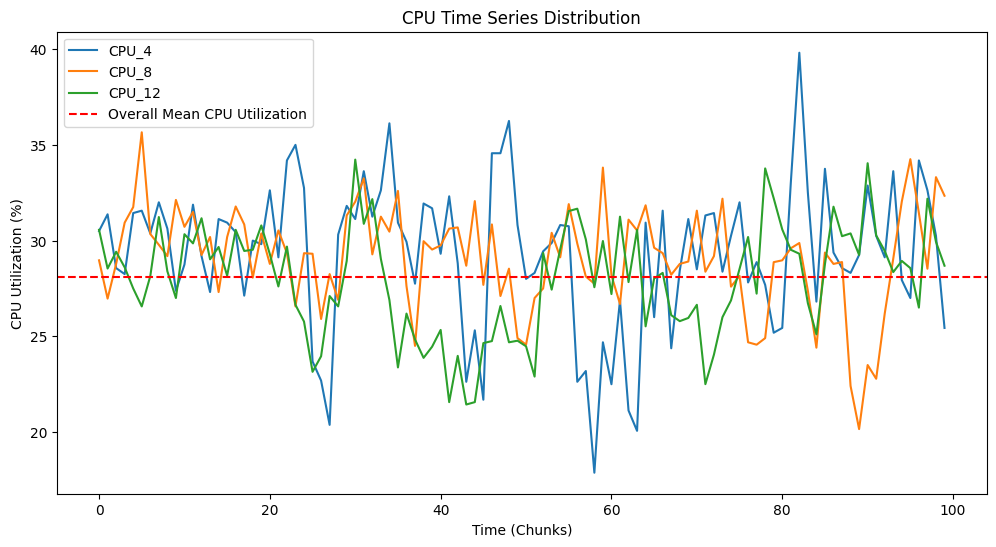}  
        \caption{CPU Utilization Time Series for Configurations}
        \label{fig:multitimecpu}
    \end{subfigure}

    \caption{Multi-Metric Analysis of RTT and CPU Utilization Across Server Setups}
    \label{fig:PerformancemultiAnalysis}
\end{figure*}

Figure \ref{fig:multimeanrtt} shows a clear trend of increasing latency with more servers. The four-server setup has the lowest median latency (about 220 ms) and the least variability. In contrast, the twelve-server setup displays a broader range (approximately 240–280 ms), highlighting the challenges of managing distributed coordination. This increased RTT variance indicates that maintaining consistent network performance becomes more complex as server count rises.

Figure \ref{fig:multimeancpu} shows that the median utilization across all setups remains stable at approximately 30\%. Notably, the 4-server configuration exhibits the most comprehensive interquartile range and shows outliers below 20\%, highlighting some inefficiencies during specific intervals. In contrast, the 8-server and 12-server setups demonstrate narrower ranges, suggesting improved load balancing, though this may come at the cost of consistency under heavier loads.

Figure \ref{fig:multitimertt} shows that latency rises as the server count increases. The four-server setup has the lowest average RTT at about 230 ms with minimal fluctuations. The eight-server setup averages around 240 ms, while the twelve-server configuration reaches approximately 270 ms. This trend highlights that additional servers lead to communication overhead and coordination delays, negatively impacting latency as setups scale.

Figure \ref{fig:multitimecpu} shows that The CPU utilization time series fluctuates around a mean value of approximately 30\% utilization for all setups. The 4-server setup exhibits higher variance with periodic spikes, suggesting an uneven load distribution. The 8-server and 12-server setups show more stable utilization but introduce occasional dips, possibly due to load redistribution inefficiencies. Including the mean CPU utilization as a reference highlights that while the system remains stable on average, transient conditions can lead to processing inefficiencies or bottlenecks.

The analysis reveals a complex relationship between Round Trip Time (RTT) and CPU utilization as server configurations scale. While increasing the number of servers reduces CPU variability and promotes more effective load balancing, it also leads to higher RTT due to increased inter-server coordination. For example, the 4-server setup exhibits higher CPU variability and frequent load spikes, whereas the 8-server and 12-server configurations demonstrate more excellent stability with occasional dips from load redistribution inefficiencies. Notably, despite similar CPU utilization levels, the 12-server setup experiences an average RTT of approximately 270 ms compared to about 230 ms for the 4-server setup. These results emphasize the trade-offs between processing efficiency and network latency, highlighting the need for careful resource balancing to optimize scalability and performance in distributed systems.

\section{Conclusion}
\label{sec:conclusions}

This paper presents dynamic algorithms and multi-metric analysis to tackle significant challenges in Content Delivery Networks (CDNs), emphasizing scalability, load balancing, and latency optimization. The study uncovers critical trade-offs between increasing server capacity and maintaining performance by examining round-trip time (RTT) and CPU utilization across various server configurations. Results obtained from FABRIC and edge environments indicate that larger server setups enhance load stability and incur greater RTT overhead due to inter-server coordination. These findings emphasize the limitations of single-metric approaches and highlight the importance of multi-metric frameworks in capturing the complexities of system dynamics. Future research will integrate additional metrics, including bandwidth and memory utilization, to enrich the analysis and provide actionable insights into network performance. Furthermore, exploring reinforcement learning models for dynamic resource management will facilitate intelligent decision-making under varying network loads. This research establishes a foundation for scalable, efficient, and adaptive CDN architectures that effectively meet real-world demands.
{\bibliographystyle{IEEEtran}}
\bibliography{paper}

\end{document}